\documentclass[twocolumn,showpacs,preprintnumbers,amsmath,amssymb]{revtex4}
\usepackage{graphicx}
\usepackage{epsfig}
\usepackage{dcolumn}
\usepackage{bm}
\begin{document}

\title{Fractal Properties of Robust Strange Nonchaotic Attractors\\
in Maps of Two or More Dimensions}
\author{Jong-Won Kim$^*$, Sang-Yoon Kim$^{\dagger}$, 
Brian Hunt$^\ddagger$, and Edward Ott$^{*,\S}$}
\affiliation{University of Maryland, College Park, Maryland 20742}
\date{\today}

\begin{abstract}
  We consider the existence of robust strange nonchaotic attractors (SNA's) 
in a simple class of quasiperiodically forced systems.
Rigorous results are presented demonstrating that the resulting attractors are 
strange in the sense that their box-counting dimension is $N+1$ while their 
information dimension is $N$. We also show how these properties are manifested
in numerical experiments.
\end{abstract}
\maketitle

\section{Introduction}

  The phrase \it strange nonchaotic attractor \rm \cite{Grebogi1} (SNA) refers
to an attractor which is nonchaotic in the sense that its orbits are not 
exponentially sensitive to perturbation (\it i.e., \rm none of the Lyapunov
exponents are positive), but the attractor is strange in the sense that
its phase space structure has nontrivial fractal properties.
Past studies indicate that SNA's are \it typical \rm in nonlinear dynamical
systems that are quasiperiodically forced. Here by a typical behavior we
mean that the behavior occurs for a positive measure set of parameter values.
Alternatively, if parameters are chosen at random from an ensemble with smooth
probability density, then the probability of choosing parameters that yield a
typical behavior is not zero. The description of a behavior as typical is to
be contrasted with the stronger statement that a behavior is \it robust.
\rm In particular, we say a behavior of a system is robust if it persists
under sufficiently small perturbations; \it i.e., \rm there exist a positive
value $\delta$ such that the robust behavior occurs for all systems that
can be obtained by perturbation of the original system by an amount less than
$\delta$. Thus all robust behaviors are also typical, but not vice versa.

  With respect to SNA's, examples where they are typical but not robust have 
been extensively studied \cite{Bondeson1,Ding1,Romeiras1,Ketoja1}. 
An example of this type is the
quasiperiodically forced circle map given by the system \cite{Ding1},
\begin{subequations}
\label{eq:map}
\begin{align}
\theta_{n+1} &= [\theta_n + \omega ] \text{~~mod $2\pi$}, \\
\varphi_{n+1} &= [\varphi_n + \omega_{\varphi} + \varepsilon \sin 
\varphi_n + C \cos \theta_n]
\text{~~mod $2\pi$},
\end{align}
\end{subequations}
where $\Omega \equiv \omega/2\pi$ is irrational. 
Other examples of typical nonrobust SNA's 
involving differential equations have also been studied 
\cite{Bondeson1,Romeiras1}. Numerical evidence \cite{Ding1,Romeiras1} 
and analysis based on a correspondence \cite{Bondeson1,Ketoja1}
with Anderson localization in a quasiperiodic potential leads to an 
understanding of the typical but nonrobust nature of SNA's in these examples:
In particular, it is found that SNA's exist on a positive Lebesgue measure
Cantor set in parameter space. In the case of Eq.~(\ref{eq:map}), for 
example, consider the rotation number
\begin{equation}
W = \lim_{n \rightarrow \infty} (\varphi_n - \varphi_0)/(2\pi n),
\end{equation}
where for this limit $\varphi_n$ is {\em not} computed modulo $2\pi$.
For fixed $\omega$, $\varepsilon > 0$, and $C>0$, a plot of $W$
versus $\omega_{\varphi}$ yields an incomplete devil's staircase, a nondecreasing
graph consisting of intervals of $\omega_{\varphi}$ where $W(\omega_{\varphi})$ is
constant and with increase of $W(\omega_{\varphi})$ occurring only on a
Cantor set of positive measure. For small $\varepsilon$, the values of 
$\omega_{\varphi}$ on the Cantor set correspond to orbits that are three
frequencies quasiperiodic, but for larger $\varepsilon$ they correspond to SNA's.
Because an arbitrarily small perturbation of $\omega_{\varphi}$ from a value
in the Cantor set can result in a value of $\omega_{\varphi}$ outside the Cantor
set, these SNA's are not robust. On the other hand, because the Cantor set of
$\omega_{\varphi}$ values has positive Lebesgue measure (`positive length'),
these attractors are typical for (\ref{eq:map}).

  Other studies suggest that there are situations where SNA's are robust
\cite{Grebogi1,Ditto1,Heagy1,Feudel1,Nishikawa1,Yalcinkaya1,Prasad1,Witt1}. 
The experiment of Ditto \it et al.\rm \cite{Ditto1} on a quasiperiodically 
forced magnetoelastic ribbon produced evidence of a SNA, and the existence of 
this SNA appeared to be stable to parameter perturbations. The original paper
where the existence of SNA's in quasiperiodically forced systems was first
discussed \cite{Grebogi1} gives numerical evidence of robust SNA's. 
In addition, the effect of quasiperiodic perturbations on a system undergoing
a periodic doubling cascade has been investigated, and evidence has been
presented suggesting that, after a finite number of torus doublings, a robust
SNA results \cite{Heagy1,Nishikawa1}.

  Thus there seems to be two types of SNA's: typical, non-robust SNA's, and
robust SNA's. In this paper we study a class of models exhibiting robust
SNA's. The model class that we study is particularly interesting because
it allows the possibility of rigorous analysis. In particular, we are able to
prove, under the mild hypothesis that a certain Lyapunov exponent is negative,
that the attractor is strange and nonchaotic. Since
other cases of SNA's are likely to be accessible only to study by numerical
means, it is worthwhile to investigate our, more well-understood models,
numerically. By doing this we gain insight into the applicability and 
limitations of numerical techniques for the study of SNA's. 

  In this paper we consider quasiperiodically forced maps which can be 
motivated by consideration of a system of ordinary differential equations in 
the form $d {\bf x}/dt = {\bf F}({\bf x},\xi,\theta^{(1)},\theta^{(2)},
\cdots,\theta^{(N)})$, where {\bf F} is $2\pi$ periodic in the angles 
$\xi$ and $\theta^{(i)}$, which are given by $\xi = \omega_{\xi}t+\xi_0$,
$\theta^{(i)} = \omega_{\theta^{(i)}}t + \theta^{(i)}_0$, and 
$\omega_{\xi}, \omega_{\theta^{(1)}}, \cdots, \omega_{\theta^{(N)}}$ 
are incommensurate. Sampling the state of the system at discrete times 
$t_n$ given by $\xi = 2n \pi$, we obtain a mapping of the form,
\begin{subequations}
\label{eq:ndmap}
\begin{align}
\theta^{(i)}_{n+1} &= [\theta^{(i)}_n + \omega^{(i)}] \text{~~mod $2\pi$},\\
{\bf x}_{n+1} &= {\bf \tilde{F}}({\bf x}_n,\theta^{(1)}_n,\theta^{(2)}_n,
\cdots,\theta^{(N)}_n),
\end{align}
\end{subequations}
where ${\bf x}_n = {\bf x}(t_n)$, 
$\omega^{(i)}=2\pi \omega_{\theta^{(i)}}/\omega_\xi$, and there exist
no set of integers $(m^{(0)}, m^{(1)}, \cdots, m^{(N)})$ for which
$\Sigma_{i=1}^N m^{(i)}\omega^{(i)} = 2\pi m^{(0)}$, aside from
$(m^{(0)}, m^{(1)}, \cdots, m^{(N)}) = (0,0,\cdots,0)$.

  For the map (\ref{eq:ndmap}), the simplest possible attractor is an
$N$-dimensional torus, ${\bf x} = {\bf f}(\theta^{(1)},\theta^{(2)},\cdots,
\theta^{(N)})$. In this paper, we consider the case where an attracting
$(N+1)$-dimensional torus exists, and the dynamics on the torus is given by
\begin{subequations}
\label{eq:ndtorus}
\begin{align}
\theta^{(i)}_{n+1} &= [\theta^{(i)}_n + \omega^{(i)}] \text{~~mod $2\pi$},\\
\varphi_{n+1} &= [\varphi_n + q^{(1)}\theta^{(1)}_n + q^{(2)}\theta^{(2)}_n+\cdots
+q^{(N)}\theta^{(N)}_n \\ \nonumber
&~~+ P(\varphi_n,\theta^{(1)}_n,\theta^{(2)}_n,
\cdots,\theta^{(N)}_n)] \text{~~mod $2\pi$},
\end{align}
\end{subequations}
where $P$ is periodic in all its variables and $q^{(1)}$, $q^{(2)}$, $\cdots$,
$q^{(N)}$ are integers. We are particularly interested in the case that
(\ref{eq:ndtorus}b) is invertible, so that no chaos is possible, and when
at least one $q^{(i)}$ is nonzero, which as we will see prevents the existence
of an attracting $N$-torus.

  In Sec.~II we examine the simplest case where $N=1$ 
($\theta^{(i)} \rightarrow \theta$).
Section~II.A presents numerical experiments and rigorous analysis of this 
two-dimensional map model. In particular, we prove (subject to a mild hypothesis
on the negativity of a Lyapunov exponent) that, for our class of maps,
the information dimension of the SNA is one ($D_1 = 1$), while its
box-counting dimension is two ($D_0 = 2$) \cite{Hunt1}. Thus we rigorously 
characterize the nature of the strangeness of the SNA's for our model.
[In a previous work \cite{Ding2} it was argued (nonrigorously) that
$D_1 =1$ and $D_0=2$ for the two-dimensional SNA map introduced in 
\cite{Grebogi1}.] We conjecture that \it $D_1 =1$ and $D_0=2$ typically holds
for SNA's of two-dimensional quasiperiodically forced maps \rm 
(\it i.e.\rm, maps of the form $\theta_{n+1} = (\theta_n + \omega)$ 
mod $2\pi$, $\varphi_{n+1} = F(\varphi_n, \theta_n)$ with 
$\Omega \equiv \omega/2\pi$ irrational). 
Also, in Sec.~II.A we present numerical experiments on dimension calculations
of $D_1$ and $D_0$, and of the Lyapunov exponent for our map. Section~II.B
investigates the dynamical origin of SNA's as a limit as $\Omega$ approaches
its irrational value through an infinite sequence of finer and finer rational
approximations (RA's) \cite{SNA1}. 
It turns out that this technique yields substantial
insight into the structure of SNA's, as well as additional understanding of
why $D_1 = 1$ and $D_0=2$ applies.

  Section~III considers higher dimensional maps. In particular, Sec.~III.A
considers the case where {\bf x} in Eq.~(\ref{eq:ndmap}b) is two-dimensional
and $N=1$, while Sec.~III.B considers $N>1$ with {\bf x} a scalar angle variable 
(as in Sec.~II). For the map of Sec.~III.A, we consider one component of
{\bf x} to be an angle variable and the other component to be analogous to
a radial variable. Thus, if, on the attractor, the radial coordinate depends
smoothly on the other two variables (which are angles), then the attractor
lies on a two-torus, and the considerations of Sec.~II apply directly.
On the other hand, the existence of such a smooth two-torus is in question,
and this is the main issue addressed in Sec.~III.A. For the map of Sec.~III.B
we are able to generalize the rigorous approach of Sec. II.A to show that
for this class of maps $D_0 = N+1$ while $D_1 = N$. In addition, numerical
experiments are performed to test the convergence of dimension computations
to these values.

\section{Two Dimensional Map on a Torus}

\subsection{Existence of SNA}

  We investigate the simplest case of (\ref{eq:ndmap}) where $N=1$
$(\theta^{(i)} \rightarrow \theta)$ and the state variable ${\bf x}$ is
one-dimensional. Specifically, we take ${\bf x}$ to be an angle variable 
$\varphi$, so that the map operates on a two-dimensional 
$\theta$-$\varphi$ torus.
Within this class we restrict consideration to maps of the form
\begin{subequations}
\label{eq:2dmap}
\begin{align}
\theta_{n+1} &= [\theta_n + \omega] \text{~~mod $2\pi$},\\
\varphi_{n+1} &= [\theta_n + \varphi_n + \eta P(\theta_n, \varphi_n) ]
\text{~~mod $2\pi$},
\end{align}
\end{subequations}
where $\omega=\pi(\sqrt{5}-1)$, and $P(\theta, \varphi)$ is continuous, 
differentiable, and $2\pi$ periodic in both of its arguments ($\theta$
and $\varphi$). When $\eta$ is small enough ($|\eta| < \eta_c$), this map is 
invertible. That is, the map is solvable for $(\theta_n, \varphi_n)$ when 
$(\theta_{n+1}, \varphi_{n+1}$) is given. We choose a simple function 
$P(\theta, \varphi) = \sin \varphi$ for our numerical work. In this case,
the system is invertible if $|\eta|<1$. Furthermore, since the map is
invariant under the change of $\eta \rightarrow -\eta$ and 
$\varphi \rightarrow \varphi + \pi$, it is sufficient to consider only the case
$\eta \ge 0$.

  Figure~\ref{fig:draw} illustrates how a curve $C$ on the $\theta$-$\varphi$ 
toroidal surface is mapped to a curve $C'$ by the map (\ref{eq:2dmap}). 
Note that the torus is unrolled in the $\theta$ direction to visualize the 
whole curve $C$ in a two-dimensional plain, but still rolled in 
the $\varphi$ direction. The curve $C$ circles around the torus 
in the $\theta$ direction, but does not wrap around the torus in the $\varphi$ 
direction. After one iterate of (\ref{eq:2dmap}), the curve $C$ is mapped to 
a curve $C'$ that wraps once around the torus in the $\varphi$ direction. 
This behavior comes about due to the term $\theta_n$ on the right-hand side of
(\ref{eq:2dmap}b), because $\theta + \varphi + \eta P(\theta, \varphi)$
increases by $2\pi$ as $\theta$ increases by $2\pi$.
Similarly, applying the map to $C'$ produces a curve with two wraps around the 
torus in the $\varphi$ direction, and so on.

  The main results of our numerical experiments and rigorous
analysis of (\ref{eq:2dmap}) with $|\eta|<\eta_c$ are as follows: 

(i) The map (\ref{eq:2dmap}) has a single attractor. 

(ii) For typical $P(\theta, \varphi)$,
the attractor has a Lyapunov exponent $h_\varphi$ that is negative for 
$\eta \neq 0$. 

(iii) The attractor has information dimension one for 
$\eta \neq 0$. 

(iv) The attractor is the entire $\theta$-$\varphi$ torus and, hence,
has box-counting dimension two \cite{Ding2}. 

(v) These results are stable to
perturbations of the system \cite{Bondeson1,Ding1,Romeiras2}. 

We first establish (ii) using an approximate formula for $h_\varphi$ 
for small $\eta$. Our evidence for (ii) is strong but a rigorous mathematical
proof is lacking. If we adopt (ii) as a hypothesis, then all the other results
rigorously follow.

  {\bf Lyapunov Exponent:} A trajectory of the map (\ref{eq:2dmap}) 
has two Lyapunov exponents $h_{\theta}$ and $h_{\varphi}$, 
where $h_{\theta} = 0$ is associated with
(\ref{eq:2dmap}a) and $h_{\varphi}$ is associated with (\ref{eq:2dmap}b).
The latter exponent is given by the formula,
\begin{equation}
h_{\varphi}= \int \ln[1+\eta P_{\varphi}(\theta,\varphi)]d\mu,
\label{eq:lyap}
\end{equation}
where $P_{\varphi} = \partial P/\partial \varphi$, and $\mu$ denotes
the measure generated by the orbit from a given initial point 
($\theta_0$, $\varphi_0$).

  If $h_{\varphi} > 0$ for a particular trajectory, then, since $h_{\theta}=0$,
the map exponentially expands areas near the trajectory in the limit
$n \rightarrow \infty$. Since the $\theta$-$\varphi$ torus has finite area, 
if the map is invertible, then there cannot be a set of
initial points of nonzero area (positive Lebesgue measure) for which 
$h_{\varphi}>0$, and the map thus does not have a chaotic attractor.
Thus $h_\varphi \le 0$ for typical orbits.

  Furthermore, we argue that  $h_{\varphi} <0$ for small nonzero $\eta$.
We consider first the case $\eta = 0$, for which (\ref{eq:2dmap}b) becomes 
$\varphi_{n+1} = (\theta_n + \varphi_n)$ mod $2 \pi$. If we initialize a uniform
distribution of orbit points in the $\theta$-$\varphi$ torus, then, on one 
application of the $\eta = 0$ map, the distribution remains uniform. 
Furthermore, this uniform distribution is generated by the orbit from
any initial condition. To verify this, we note that the explicit form
of an $\eta = 0$ orbit, $\theta_n = (\theta_0 + n \omega)$ mod $2\pi$,
$\varphi_n = [\varphi_0 + n \theta_0 + \frac{1}{2} (n^2 - n)\omega]$ 
mod $2\pi$, which
is shown to generated a uniform density in Ref. \cite{Furstenberg1}. 
We can obtain an approximation to $h_\varphi$ for nonzero but small $\eta$ by 
expanding $\ln (1+\eta P_\varphi)$ in (\ref{eq:lyap}) to order $\eta^2$ and 
assuming that, to this order, the deviation of the measure $\mu$ from
uniformity is not significant $[d\mu \approx d\theta d\varphi /(2\pi)^2]$. Using 
$\ln(1+\eta P_\varphi) = \eta P_\varphi - (1/2)\eta^2 P^2_\varphi + O(\eta^3)$, 
this gives
\begin{equation}
h_\varphi = -\frac{1}{2}\eta^2<P^2_\varphi> + o(\eta^2),
\label{eq:hlyp}
\end{equation}
which is negative for small enough $\eta \neq 0$. Here $<P^2_\varphi>$
denotes the $\theta$-$\varphi$ average of $P^2_\varphi$, and the order $\eta$
term is absent by virtue of $\int_0^{2\pi} P_\varphi d\varphi=0$.
Since we cannot show convergence of an expansion in $\eta$, our result
(\ref{eq:hlyp}) is formal rather than rigorous. However, numerical results
strongly support (\ref{eq:hlyp}).
Figure~\ref{fig:lyap} shows a plot of $h_\varphi$ versus $\eta$ for
$P(\theta,\varphi) = \sin \varphi$. Remarkably, Eq.~(\ref{eq:hlyp}) 
(the straight line) describes the numerical data to better than 8 \% even for
$\eta$ as large as 0.5.

  {\bf Dimensions of the SNA:}
For our map the information dimension cannot be less than one due to 
the quasiperiodic $\theta$ dynamics. In addition, the Lyapunov dimension is an 
upper bound of information dimension \cite{Ledrappier1}. Therefore, if we 
accept (ii), $h_\varphi < 0$, then $h_\theta = 0$ implies (iii). 

  Results (iii) and (iv) quantify the strangeness of the attractor. 
In particular, since the information dimension of the attractor is one, 
orbits spend most of their time on a curve-like set;
yet, since the box-counting dimension is two, if one waits long enough,
a typical orbit eventually visits any
neighborhood on the $\theta$-$\varphi$ torus. One can get a sense of this result
from the numerical orbit shown in Fig.~\ref{fig:traj2}, in which a trajectory 
of length $10^4$ appears to be concentrated along one-dimensional strands
[Fig.~\ref{fig:traj2}(a)], but for the
same parameters a trajectory of length $10^5$ fills much more of the
$\theta$-$\varphi$ torus [Fig.~\ref{fig:traj2}(b)].

  We show in Fig.~\ref{fig:bid2}(a) a plot of $\log_2 N(\varepsilon)$ versus
$\log_2 (1/\varepsilon)$, and in Fig.~\ref{fig:bid2}(b) a plot of 
$\Sigma p_i \log_2 (1/p_i)$ versus $\log_2(1/\varepsilon)$. 
Here $N(\varepsilon)$ is the number of $\varepsilon \times \varepsilon$ 
boxes (in $\theta$-$\varphi$ space) needed to cover the points from an orbit of 
length $T$, and $p_i$ is the fraction of those orbit points in the 
$i$th $\varepsilon \times \varepsilon$ box.
According to our previous arguments on dimensions, in the limit 
$T \rightarrow \infty$, the points in Fig.~\ref{fig:bid2}(a) and 
Fig.~\ref{fig:bid2}(b) should follow a straight line of slope two and one for 
small $\varepsilon$, corresponding to a box-counting dimension of two and
an information dimension of one. As is commonly found, the box-counting
dimension computation converge rather slowly with increasing orbit length
$T$. Thus, we show plots in Fig.~\ref{fig:bid2} for several different $T$.
As can be seen in Fig.~\ref{fig:bid2}(a), the $\varepsilon$ range consistent
with a slope of two (the straight line in the figure) steadily increases
toward smaller $\varepsilon$ [larger $\log(1/\varepsilon)$] as $T$ increases.
This is contrast with Fig.~\ref{fig:bid2}(b), which appears to reach a 
form independent of $T$ that is consistent with a small $\varepsilon$ slope
of one. While the convergence in Fig.~\ref{fig:bid2}(a) and \ref{fig:bid2}(b)
is consistent with box-counting and information dimensions of two and one,
the slowness of the convergence also indicates that a purely numerical 
determination of the dimension values is suspect.

  {\bf Topological Transitivity:}
  To establish results (i) and (iv), that the attractor of the map is
 in the whole $\theta$-$\varphi$ torus, 
we prove that the map is \it topologically transitive: \rm
For every pair of open disks $A$ and $B$, there is a trajectory that starts
in $A$ and passes through $B$. This property is known to imply that a
dense set of initial conditions yields trajectories each of which is dense in
the torus \cite{Katok1}. In particular, any attractor, having an open
basin of attraction, must contain a dense orbit, and, hence, must be the
entire torus.

  We will show in fact that for every pair of line segments
$S_a = \{(\theta,\varphi): \theta \in R_a$ and $\varphi = \varphi_a \}$ and
$S_b = \{(\theta,\varphi): \theta \in R_b$ and $\varphi = \varphi_b \}$, where
$R_a = (\theta_a, \theta_a + \delta_a)$ and
$R_b = (\theta_b, \theta_b + \delta_b)$, there is a finite trajectory of $M$
that begins on the first segment and ends on the second. (Choosing $S_a$ to lie
in $A$ and $S_b$ to lie in $B$, this implies topological transitivity.) 
In other word, we will show that the $n$th iterate of $S_a$ intersects $S_b$
for some positive integer $n$; see Fig.~\ref{fig:dia2}(a). Our strategy is to
iterate $S_a$ forward until the union of its iterates covers all values
of $\theta$ at least once; the number of iterates needed is finite and depends 
only on $\delta_a$. By selecting pieces of some of these iterates that cover
each value of $\theta$ exactly once, we form the graph $\varphi = g_a(\theta)$ 
of a piecewise continuous function $g_a$; see Fig.~\ref{fig:dia2}(b).
Similarly we form a graph $\varphi = g_b(\theta)$ from pieces of backward 
iterates of $S_b$. Finally, we show that some forward iterate of the graph of 
$g_a$ must intersect the graph of $g_b$.

  The following is a formal definition of $g_a$. Let $M_\theta$ be the map
(\ref{eq:2dmap}a). For each $\theta$, let $k(\theta)$ be the smallest 
nonnegative integer for which $\theta \in M^k_\theta (R_a)$. 
(In Fig.~\ref{fig:dia2}(b), $k(\theta)=$ 0, 1, or 2 for all $\theta$.)
Let $g_a(\theta)$ be the $\varphi$-coordinate of the $k(\theta)$-th iterate
under $M$ of $(M^{-k(\theta)}_\theta, \varphi) \in S_a$. Then the graph
$\varphi = g_a(\theta)$ has a finite number of $d_a$ of discontinuities.
Each contiguous piece of this graph is a forward iterate of some piece
of $S_a$.

  Now form the curve $G_a$ by taking the graph of $g_a$ and adding
line segments in the $\varphi$ direction at each value of $\theta$ where
$g_a$ is discontinuous. (We take these segments to lie in $0<\varphi<2\pi$.)
Thus we make $G_a$ a contiguous curve. See 
Fig.~\ref{fig:dia2}(b), which illustrates this construction on for a case where
$d_a = 3$.  Notice that, for
each $n$, the $n$th iterate of $G_a$ is also a contiguous curve that
consists of the graph of a function with $d_a$ discontinuities, together with 
$d_a$ ``connecting segments". Define $g_b$ and $G_b$ similarly to 
$g_a$ and $G_a$, but in terms of the backward (not forward) iterates of the 
$S_b$. Let $d_b$ be the number of discontinuities of $g_b$. 

  Our goal is to show that for $n$ sufficiently large, the $n$th iterate
of $G_a$ intersects $G_b$ for at least $d_a + d_b +1$ different values of
$\theta$.  Then since there are at most $d_a + d_b$ values of
$\theta$ at which one of these two curves has a connecting segment,
there will be at least one intersection point between the $n$th iterate of the
graph of $g_a$ and the graph of $g_b$. Since the graph of $g_a$ consists of
forward iterates of $S_a$ and the graph of $g_b$ consists of backward 
iterates of $S_b$, some forward iterated $S_a$ will intersect $S_b$,
as we claimed.

  Given a contiguous curve $C$ that, like $G_a$ and $G_b$, is the graph
of a function of $\theta$ that is continuous except for a finite number of 
values at which $C$ has a connecting segment, observe that its image under
$M_\theta$ is a curve of the same type (in particular, since the
map is one-to-one, the heights of the vertical segments remain less than 
$2\pi$).  Furthermore, because of the $\theta_n$ term
in the $\varphi$ map (\ref{eq:2dmap}b), the image of $C$ ``wraps around''
the torus in the $\varphi$ direction one more time than $C$ does as
one goes around the torus one time in the $\theta$ direction 
(see Fig.~\ref{fig:draw}).

  To formulate what we mean by ``wrapping around'', define the winding
number of $C$ as follows.  As $\theta$ increases from $0$ to $2\pi$, 
count the number of times $C$ crosses $\varphi = 0$ in the upward and downward 
directions. The difference between the number of upward and downward crossings 
is the winding number of $C$. (The numbers of upward and downward
crossings may depend on the arbitrary choice of 0 as the $\varphi$ value at
which to count crossings, but their difference does not.) For example,
in Fig.~\ref{fig:dia2}(b) the winding number of $G_a$ is 0.

  Now if two curves $C_1$ and $C_2$, as described above, have different
winding numbers $w_1$ and $w_2$, then $C_1$ and $C_2$ must intersect
at least $|w_1 - w_2|$ times. Because of the periodicity of 
$P(\theta, \varphi)$, the winding number of a curve must increase by 1
each time the map $M$ is applied.  Thus for $n$ sufficiently large, 
the winding number of the $n$th iterate of $G_a$
differs from the winding number of $G_b$ by at least $d_a + d_b + 1$.
Hence the $n$th iterate of $G_a$ intersects $G_b$ for at least 
$d_a + d_b + 1$ different values of $\theta$ as desired. 
This establishes claims (i) and (iv).

  Notice that the argument above does not depend on the specific form
of $P(\theta,\phi)$, only that it is continuous and periodic and that
$\eta$ is sufficiently small $(|\eta|<\eta_c)$ that the map
(\ref{eq:2dmap}) is one-to-one. This independence of the results from
the specific form of $P(\theta,\phi)$ implies that the results are stable
to system changes [our claim (v)] that preserve a quasiperiodic driving
component (\ref{eq:2dmap}a).

  {\bf Discussion:} The possible existence of SNA's was originally pointed
out in \cite{Grebogi1}, and many numerical explorations of the dynamics on
attractors that are apparently strange and nonchaotic have appeared.
Recently, there has also been rigorous results on the mathematical properties
that SNA's must have if they exist \cite{Stark1}. In spite of these works,
a very basic question has remained unanswered: {\it Can it be rigorously
established that SNA's generically exist in typical quasiperiodically forced
systems?} This is an important issue, because, although the numerical evidence
for SNA's is very strong, perhaps the attractors observed are nonstrange
with very fine scale structure (rather than the {\it infinitesimally} fine 
scale structure of a truly strange attractor). Also, there might be the
worry that the numerical evidence is somehow an artifact of computational
error. Our proof of topological transitivity, combined with the hypothesis that
$h_\varphi < 0$, answers the question of the typical existence of SNA's 
(affirmatively) for the first time (\cite{Hunt1} contains a preliminary report 
of our work). The only previous work rigorously establishing the existence of 
a SNA is that appearing in the original publication on SNA's \cite{Grebogi1}
and in \cite{Keller1}. These proofs, however, are for a very special class
of quasiperiodically forced system such that an arbitrarily small typical
change of the system puts it out of the class. Thus this proof does not 
establish that SNA's exist in typical quasiperiodically forced situations.
In order to see that nature of this situation with respect to Refs.
\cite{Grebogi1} and \cite{Keller1}, we recall the example treated in 
Ref. \cite{Grebogi1}. In that reference the map considered was
$x_{n+1} = 2 \lambda (\tanh x_n) \cos \theta_n \equiv f(x_n, \theta_n)$,
with $\theta_n$ evolved as in Eq.~(\ref{eq:2dmap}a). It was proven in
\cite{Grebogi1} that this map has a SNA for $\lambda > 1$.
However, the map has an invariant set, namely, the line $x=0$, 
$\theta$ in $[0, 2\pi)$, and this fact is essential in the proof of
Ref. \cite{Grebogi1}. On the other hand, the existence of this invariant
set does not persist under perturbations of the map. Thus, if we perturb
$f(x,\theta)$ to $f(x,\theta)+\varepsilon g(x,\theta)$, the invariant set is
destroyed, even for small $\varepsilon$, for any typical function $g(x,\theta)$
(in particular, an arbitrarily chosen $g(x,\theta)$ is not expected to
satisfy $g(0,\theta)=0$).

\subsection{Origin of SNA's: Rational Approximation}

  Using rational approximations (RA's) to the quasiperiodic forcing,
we now investigate the origin for the appearance of SNA's in
(\ref{eq:2dmap}) for $P(\theta, \varphi)=\sin \varphi$ and 
$\omega = \pi(\sqrt{5}-1)$. For the case of the inverse golden mean
$\Omega \equiv \omega/2\pi$,
its rational approximants are given by the ratios of the Fibonacci
numbers, $\Omega_k = F_{k-1} / F_k$, where the sequence of $\{ F_k
\}$ satisfies $F_{k+1} = F_k + F_{k-1}$ with $F_0 = 0$ and $F_1 =
1$. Instead of the quasiperiodically forced system, we study an
infinite sequence of periodically forced systems with rational
driving frequencies $\omega_k$. We suppose that the properties of
the original system may be obtained by taking the
quasiperiodic limit $k \rightarrow \infty$. 

  For each RA of level $k$, a periodically forced map with the
rational driving frequency $\Omega_k$ has a periodic or
quasiperiodic attractor that depends on the initial phase
$\theta_0$ of the external force. Then we take the union of all attractors
for different $\theta_0$ to be the $k$th RA to the attractor in
the quasiperiodically forced system. Furthermore, due to the
periodicity, it is sufficient to obtain the RA by changing
$\theta_0$ only in an basic interval $\theta_0 \in [0, 1/F_k)$,
because the RA to the attractor in the remaining range, $[1/F_k,1)$, may be 
obtained through $(F_k - 1)$-times iterations of the result in $[0,1/F_k)$.
For a given $k$ we call the periodic attractors of period $F_k$ the ``main
periodic component". The first column of Fig.~\ref{fig:ra1} shows that the 
Lebesgue measure of 
the main periodic component (denoted by the solid line) becomes
dominant as the level $k$ increases (\it i.e.\rm, the fraction of the $\theta$
axis corresponding to the non-periodic, gray area decreases). 
By iterating the RA in the
basic interval of $\theta$, we obtain the RA in the whole range of
$\theta$, as shown in the second column of Fig.~\ref{fig:ra1}. As
$k$ increases, the whole RA becomes more similar to its
quasiperiodic limit given in Fig.~\ref{fig:traj2}. 

We first note that for $\eta=0$ the RA to the regular
quasiperiodic attractor consists of only the quasiperiodic
component. However, as $\eta$ becomes positive periodic components
appear via phase-dependent (\it i.e.\rm, $\theta_0$-dependent)
saddle-node bifurcations. As an
example, we show the $6$th RA for $\eta=0.3$ in Fig.~\ref{fig:ra2}. 
Here the quasiperiodic component is plotted in the gray. 
``Gaps'' in the gray quasiperiodic regions are occupied by periodic attractors.
As examples, we explicitly
show the main period-$F_6$ $(F_6=8)$ and minor period-$3F_6$
components in Figs.~\ref{fig:ra2}(a) and \ref{fig:ra2}(b),
respectively. At both ends of each gap, a pair of stable (denoted
by a solid line) and unstable (denoted by a dashed line) periodic
orbits appear via a phase-dependent saddle-node bifurcation.
Figures~\ref{fig:ra3}(a) and \ref{fig:ra3}(b) show the saddle-node
bifurcation curves in the $\theta F_k$-$\eta$ plane, at which the
main periodic components with period $F_k$ are born. It can be
easily seen that for a given $\eta$ the width of the main gap
(occupied by a period-$F_k$ attractor) becomes larger as $k$ increases.
Quantitatively, it is found that the Lebesgue measure $\mu_k$ in
$\theta$ for the main periodic component becomes dominant as $k$
increases; \it i.e.\rm, the Lebesgue measure $(1 - \mu_k)$ of the
complementary set decreases exponentially with $F_k$;
$1 - \mu_k \sim e^{- \alpha F_k}$, where $\alpha=0.013$, as
shown in Fig.~\ref{fig:ra3}(c) for $\eta=0.3$.

  In what follows we use the RA's to explain the origin of
the negative Lyapunov exponent $h_\varphi$ and the strangeness of the SNA.
For a given level $k$ of the RA, let $h^{(k)}_\varphi (\theta)$ denote
the Lyapunov exponent of the attractor corresponding to a given $\theta$.
Thus $h^{(k)}_\varphi (\theta) = 0$ for $\theta$ in the quasiperiodic
range (gray regions of Figs.~\ref{fig:ra1} and \ref{fig:ra2}) and
$h^{(k)}_\varphi (\theta) < 0$ for $\theta$ in the periodic range (gaps in
the gray regions).
Since the attractor with irrational $\omega$ generates a uniform density in
$\theta$ [see Eq.~(\ref{eq:2dmap}a)], we take the order-$k$ RA to the
Lyapunov exponent $h_\varphi$ to be
\begin{equation} 
<h_\varphi^{(k)}> = \frac{1}{2\pi}\int_0^{2\pi}h_\varphi^{(k)}(\theta)
d \theta .
\end{equation}

   For $\eta>0$, due to the existence of periodic
components, $<h_\varphi^{(k)}>$ is negative. 
As $\eta$ increases for a given level $k$, the Lebesgue measure in 
$\theta$ for the periodic components increases, and 
hence $h^{(k)}_\varphi(\theta)$ becomes negative in a
wider range in $\theta$, as shown in Figs.~\ref{fig:ra4}(a) and
\ref{fig:ra4}(b) for level $k=6$. Thus, as $\eta$ increases 
$<h^{(k)}_\varphi>$ decreases [see Fig.~\ref{fig:ra4}(c)].

  In addition, we note that as the level $k$ increases, the 
RA to the Lyapunov exponent $h_\varphi$ converges rapidly to its
quasiperiodic limit [represented by the solid line in Fig.~\ref{fig:ra4}(c)]. 
For comparison, the approximate analytic result for $h_\varphi$
(i.e., $h_\varphi = - \eta^2/4$) is also given (see the dashed
line). Consequently, for any nonzero $\eta$ the attractor in the
quasiperiodic limit has a Lyapunov exponent whose value decreases
as $\eta$ is increased. 

  We now discuss the strangeness of the
attractor in the quasiperiodic limit for $\eta=0.3$. 
In the quasiperiodic case, we have seen (Fig.~\ref{fig:traj2}) that 
a typical trajectory seems to fill the whole torus densely, 
but, unlike the case of the regular quasiperiodic attractor, it appears 
to spend most of its time on a set of 1D strands.
We identify these apparent 1D strands with the
$k \rightarrow \infty$ limit of the main periodic component. Although, as
$k$ becomes larger, the Lebesgue measure of the quasiperiodic region
approaches zero [Fig.~\ref{fig:ra3}(c)], these quasiperiodic region
become dense in $\theta$. Since each quasiperiodic region fully
covers the $\varphi$ interval $[0, 2\pi)$, the attractor is expected to occupy
the entire $\theta$-$\varphi$ torus, and, hence, it is expected to have
a box-counting dimension of 2. 

\section{High Dimensional Maps}

\subsection{Radial Perturbations of the Torus Map}

  We now show that stability to perturbations applies in addition if the 
system is higher dimensional. In particular, we discuss the case of a 
three-dimensional system with an attracting invariant torus, and
allow perturbations of the toroidal surface. Consider the following
map on ${\bf R}^3$:
\begin{subequations}
\label{eq:rdmap}
\begin{align}
\theta_{n+1} &= [\theta_n + \omega] \text{~~mod $2\pi$},\\
\varphi_{n+1} &= [\theta_n + \varphi_n + \eta \bar{P}(\theta_n,\varphi_n,r_n) ]
\text{~~mod $2\pi$}, \\
r_{n+1} &= \lambda r_n + \rho Q(\theta_n,\varphi_n,r_n).
\end{align}
\end{subequations}
Here $\theta$ and $\varphi$ are coordinates on a torus embedded in ${\bf R}^3$, 
as in Fig.~\ref{fig:draw}, and $r$ is a coordinate transverse to the torus, 
with $r=0$ representing the unperturbed ($\lambda = \rho = 0$) torus.
The parameters $\omega$ and $\eta$, and the dependence of $\bar{P}$ on
$\theta$ and $\varphi$, have the same properties as for map (\ref{eq:2dmap}),
and $Q$ is continuously differentiable and $2\pi$ periodic in $\theta$
and $\varphi$. When $\lambda$ and $\rho$ are small,
Eq.~(\ref{eq:rdmap}) maps a neighborhood of the torus $r=0$ into itself, and
when $\rho=0$ the torus $r=0$ is invariant and attracting. It then follows
from classical results on the perturbation of invariant manifolds
\cite{Katok1} that, for $\lambda$ and $\rho$ sufficiently small, the map
(\ref{eq:rdmap}) has a smooth attracting invariant manifold 
$r=f(\theta,\varphi)$ near the torus $r=0$. On this attractor, the map
(\ref{eq:rdmap}) reduces to a map of the form (\ref{eq:2dmap}), with
$P(\theta,\varphi)=\bar{P}[\theta,\varphi,f(\theta,\varphi)]$. Thus
statements (i)-(v) above apply also to the attractor of the three-dimensional
map (\ref{eq:rdmap}).

  The above arguments depend on the existence of a smooth invariant torus
on which the attractor is located, and this is guaranteed if $\lambda$ and
$\rho$ are sufficiently small. We now show numerical evidence for the 
existence of a smooth invariant torus for values of $\lambda$ and $\rho$
that are appreciable. We consider the example 
$\bar{P}(\theta, \varphi, r) = \sin \varphi$ and 
$Q(\theta, \varphi, r) = \sin (r+\varphi)$.
We numerically obtain two-dimensional plots of intersections of the
invariant torus with the surfaces $\varphi = \pi$ [Fig.~\ref{fig:att}(a)]
and $\varphi = \pi$ [Fig.~\ref{fig:att}(b)].
First consider the case $\varphi = \pi$. Our numerical technique is as follows.
We choose an initial value $(\theta_0, \varphi_0=\pi)$ and obtain
$(\theta_{-n}, \varphi_{-n})$ by iterating 
(\ref{eq:rdmap}a) and (\ref{eq:rdmap}b) backward $n$ steps.
Since $h_r \sim \ln \lambda < 0$ (when $\rho \ll \lambda < 1$), 
$r_{-n} \rightarrow \pm \infty$ if $r_0$ is not on the torus. In other words, if
$r_{-n} = 0$, then $r_0$ is on the torus. Thus, we choose $r_{-n} = 0$ and 
iterate $(r_{-n}, \theta_{-n}, \varphi_{-n})$ forward
$n$ steps to $(r_0, \theta_0, \varphi_0 = \pi)$.
By varying $\theta_0$, we obtain the graph, $r_0(\theta_0)$, of the torus 
intersection with $\varphi=\pi$. 
Similarly, choosing $(\theta_0=\pi, \varphi_0)$ and iterating the map
(\ref{eq:rdmap}) backward, and then forward,
we can obtain $r_0(\varphi_0)$ of the torus intersection with $\theta=\pi$.
(For our numerical experiments, we set $n = 25$.) As shown in Fig.~\ref{fig:att},
a smooth invariant torus exists in the parameter region where $\rho$ and $\lambda$
are appreciable. 

  In Figs.~\ref{fig:bidr}(a) and (b) we show dimension computations for 
the map (\ref{eq:rdmap}) with
$\bar{P}(\theta, \varphi, r)=\sin(\varphi)$ and 
$Q(\theta, \varphi, r)=\sin(r+\varphi)$. [In this three dimensional 
case we employ $\varepsilon$ edge length cubes in 
($\theta, \varphi, r$)-space.] 
As for Fig.~\ref{fig:bid2}, we observe from Fig.~\ref{fig:bidr}, that the 
results are consistent with slow convergence to the predicted dimension 
values of 2 and 1 as the orbit length $T$ is increased.
 
\subsection{Map on a High Dimensional Torus}

  In Section~II.A we proved that (\ref{eq:2dmap}) is topologically transitive. 
Here we show how this argument can be modified to higher dimensional maps
that include $N > 1$ quasiperiodic driving variables 
$\theta^{(1)}$, $\theta^{(2)}$, $\cdots$, $\theta^{(N)}$.
For exposition we assume $N = 2$, but the argument is virtually
identical for all $N$. 

  In particular, we consider a map of the form,
\begin{subequations}
\label{eq:3dmap}
\begin{align}
\theta^{(1)}_{n+1} &= [\theta^{(1)}_n + \omega^{(1)}] \text{~~mod $2\pi$},\\
\theta^{(2)}_{n+1} &= [\theta^{(2)}_n + \omega^{(2)}] \text{~~mod $2\pi$},\\
\varphi_{n+1} &= [q^{(1)}\theta^{(1)}_n + q^{(2)}\theta^{(2)}_n + \varphi_n +\\
\nonumber 
& ~~~\eta P(\theta^{(1)}_n, \theta^{(2)}_n, \varphi_n) ] \text{~~mod $2\pi$},
\end{align}
\end{subequations}
where $\omega^{(1)}$ and $\omega^{(2)}$ are incommensurate, 
$(q^{(1)}, q^{(2)})$ is a pair of integers different from $(0,0)$, and
$P(\theta^{(1)}, \theta^{(2)}, \varphi)$ is continuous, differentiable, and
$2\pi$ periodic in all of its arguments ($\theta^{(1)}$, $\theta^{(2)}$,
and $\varphi$). We assume without loss of generality that $q^{(1)} \neq 0$. 

  Let $R_a = \{(\theta^{(1)}, \theta^{(2)}) : \theta^{(1)}_a < \theta^{(1)} <
(\theta^{(1)}_a + \delta_a) \text{ and } \theta^{(2)}_a < \theta^{(2)} <
(\theta^{(2)}_a + \delta_a)\}$ and $R_b = \{(\theta^{(1)}, \theta^{(2)}) :
\theta^{(1)}_b < \theta^{(1)} < (\theta^{(1)}_b + \delta_b) \text{ and }
\theta^{(2)}_b < \theta^{(2)} < (\theta^{(2)}_b + \delta_b)\}$ be two 
arbitrary
squares in the $\theta^{(1)}$-$\theta^{(2)}$ torus, and let $S_a =
\{(\theta^{(1)}, \theta^{(2)}, \varphi) : (\theta^{(1)}, \theta^{(2)}) \in R_a
\text{ and } \varphi = \varphi_a\}$ and $S_b = \{(\theta^{(1)},
\theta^{(2)}, \varphi) : (\theta^{(1)}, \theta^{(2)}) \in R_b \text{ and }
\varphi = \varphi_b\}$ be a pair of square segments, 
where $\varphi_a$ and $\varphi_b$ are
arbitrary.  As before, we will show that there is a finite trajectory
that begins on $S_a$ and ends on $S_b$.

In this case, we proceed by iterating $R_a$ forward until the union of
its iterates cover all points $(\theta^{(1)}, \pi)$ at least once (see
Fig.~\ref{fig:dia3}). The number of iterates needed is finite. 
Then we select pieces of these iterates that
single-cover a thin strip $D_a = \{(\theta^{(1)},
\theta^{(2)}) : \pi \leq \theta^{(2)} \leq \pi + \varepsilon_a\}$ with
rectangles of width $\varepsilon_a$.  From the corresponding pieces of
the corresponding iterates of $S_a$, we form the graph $\varphi =
g_a(\theta^{(1)}, \theta^{(2)})$ of a piecewise continuous function $g_a$
defined on $D_a$.  Similarly we form a graph $\varphi =
g_b(\theta^{(1)}, \theta^{(2)})$ on a strip $D_b$ from
pieces of backward iterates of $R_b \times \{\varphi_b\}$.  As before,
we will show that some forward iterate of the graph of $g_a$ must
intersect the graph of $g_b$.

  Next, form the strip $G_a$ by taking the graph of $g_a$ and adding
``connecting faces'' at each of the $d_a$ values of $\theta^{(1)}$ where
$g_a$ is discontinuous, so as to make $G_a$ a contiguous strip.  The
construction of $G_a$ is essentially as shown in Fig.~\ref{fig:dia2}(b), 
except that it now has some thickness in the $\theta^{(2)}$ direction 
(not shown). For each $n$, the $n$th iterate of $G_a$ is also a
contiguous strip that consists of the graph of a function with $d_a$
discontinuities in the $\theta^{(1)}$ direction, together with $d_a$
connecting faces, over a strip in the $\theta^{(1)}$-$\theta^{(2)}$ torus
of width $\varepsilon_a$ in the $\theta^{(2)}$ direction.  Notice though
that the strip moves a distance $\omega^{(2)}$ in the $\theta^{(2)}$
direction with each iteration.  Define $g_b$ and $G_b$ similarly to
$g_a$ and $G_a$, but in terms of the backward iterates of $S_b$, and
let $d_b$ be the number of values of $\theta^{(1)}$ at which $g_b$ is
discontinuous.

  As before, we can define the winding number of strips like $G_a$ and
$G_b$, representing the net number of times the strip wraps in the
$\varphi$ direction as $\theta^{(1)}$ increases from $0$ to $2\pi$.  The
winding number can be computed for any fixed value of $\theta^{(2)}$ and
does not depend on that value. With each iteration of (\ref{eq:3dmap}), the
winding number of such a strip changes by $q^{(1)} \neq 0$.  Therefore
for $n$ sufficiently large, the winding number of the $n$th iterate of
$G_a$ differs from the winding number of $G_b$ by at least $d_a + d_b
+ 1$.  Furthermore, by increasing $n$ if necessary, we can ensure that
the domains of these two strips intersect; that is, they have a common
value of $\theta^{(2)}$.  Then for that value of $\theta^{(2)}$, it
follows as before that the $n$th iterate of the graph of $g_a$
(without the $d_a$ connecting faces of $G_a$) and the graph of $g_b$
(without the $d_b$ connecting faces of $G_b$) must intersect as
claimed.

  Our numerical experiments also give a sense of the above proof.
In order to obtain two-dimensional plots, we count points when the trajectory  
passes through a thin slab of width $\delta \ll 1$ containing the 
$\theta^{(i)}$-$\varphi$ surface. (For our 
experiments, the width of the slab is $\delta =0.01$.)
Figures~\ref{fig:traj3} show numerical approximations, thus, obtained, to the
intersections of the attractors with the $\theta^{(2)}=\pi$ surface
[Fig.~\ref{fig:traj3}(a)] and the $\theta^{(1)}=\pi$ surface
[Fig.~\ref{fig:traj3}(b)]. Both figures look similar to the figure
for two dimensional case (Fig.~\ref{fig:traj2}).
We also show in Fig.~\ref{fig:bid3}(a) a plot of log$_2N(\varepsilon)$ versus
log$_2(1/\varepsilon)$, and in Fig.~\ref{fig:bid3}(b) a plot of
$\Sigma p_i \log_2(1/p_i)$ versus log$_2(1/\varepsilon)$. According to the
above proof, the box-counting dimension should be three and the information
dimension is two. However, due to the slowness of the convergence of slopes,
the dimension values are not clearly evident, although, as in the 
two-dimensional case (Sec. II.A) the results are suggestive.

\section{Conclusion}

  In this paper we addressed the existence of robust strange nonchaotic
attractors. In particular, we provided rigorous analysis for the 
two-dimensional map (\ref{eq:2dmap}) in Sec.~II.A and for the 
$(N+1)$-dimensional maps of the form of (\ref{eq:ndmap}) in Sec.~III.B. 
In addition, we have used a rational approximation technique (Sec.~II.B) 
to investigate the dynamical origin of SNA's, as well as to gain additional 
understanding on why $D_1 = 1$ and $D_0 = 2$. In Sec.~III.A, we show that
the stability to perturbations (robustness) continues to apply in systems
[\it e.g.\rm, (\ref{eq:rdmap})] where there can be an attracting torus.
We also carried out calculations to see how our rigorous dimension
results are manifested numerically. Our results confirm the existence of
SNA's as a generic phenomenon of quasiperiodically forced systems. 

\begin{acknowledgments}
S.Y.K. thanks W. Lim for his help in the numerical work. This work was
supported by the Korea Research Foundation (Grant No. KRF-2001-013-D00014), 
by the ONR (Physics) and by the NSF (DMS-0104087 and PHYS-0098632).
\end{acknowledgments}

$^*$Institute for Research in Electronics and Applied\\
\text{~~~~}Physics, and Department of Physics.

$^\dagger$Institute for Research in Electronics and Applied\\
\text{~~~~}Physics. Permanent address: Department of Physics,\\
\text{~~~~}Kangwon National University, Chunchon,\\
\text{~~~~}Kangwon-Do 200-701, Korea.

$^\ddagger$Institute for Physical Science and Technology,\\
\text{~~~~}and Department of Mathematics.

$^\S$Department of Electrical and Computer Engineering.

\newpage

\begin{figure}[p]
\caption{Torus unwrapped in the $\theta$ direction ($\theta = 0$ and
$\theta=2\pi$ are identified with each other). The map (\ref{eq:2dmap})
takes the curve $C$ to the curve $C'$.}
\label{fig:draw}
\end{figure}

\begin{figure}[p]
\caption{Lyapunov exponent $h_{\varphi}$ versus $\eta^2$. For each $\eta$, the
data plotted as open circles were computed from $10^7$ iterations of the
map (\ref{eq:2dmap}) with $\omega= \pi(\sqrt{5}-1)$ and 
$P(\theta,\varphi)= \sin \varphi$.}
\label{fig:lyap}
\end{figure}

\begin{figure}[p]
\caption{Trajectory of the map (\ref{eq:2dmap}) with $\omega=\pi(\sqrt{5}-1)$,
$\eta = 0.3$, and $P(\theta,\varphi)= \sin \varphi$. In each case 
$\theta_0 = \varphi_0 = 0$ and $10^4$ points of the trajectory are computed
before plotting; in (a) the next $10^4$ points are plotted, while
(b) shows $10^5$ points.}
\label{fig:traj2}
\end{figure}

\begin{figure}[p]
\caption{Dimension computations for (\ref{eq:2dmap}) with $\eta = 0.3$,
$\omega=\pi(\sqrt{5}-1)$, and $P(\theta,\varphi)=\sin \varphi$. In (a) the
dashed line has slope two, while in (b) it has slope one. In each graph,
the curves from lowest to highest represent $T=10^3,10^4,\cdots,10^{10}$;
in (b) the final five curves are virtually identical.}
\label{fig:bid2}
\end{figure}

\begin{figure}[p]
\caption{(a) The $n$th iterate of $S_a$ intersects $S_b$. The $n$th pre-iterate
of this intersection point (denoted $p$) is a point on $S_a$ that goes to
$S_b$ in $n$ iterates. (b) $S_a$ plus its first two iterates, $M(S_a)$ and
$M^2(S_a)$, cover the entire $2\pi$ range of $\theta$. $M(S_a)$ and $M^2(S_a)$
are shown as thin lines. The curve $G_a$, which includes $S_a$, 
pieces of $M(S_a)$ and $M^2(S_a)$, and vertical segments connecting these
pieces, is shown as a dark thick line.}
\label{fig:dia2}
\end{figure}

\begin{figure}[p]
\caption{RA's for $\eta=0.3$. The levels are $k=6$ in (a) and (b),
$k=8$ in (c) and (d), and $k=11$ in (e) and (f). In the first
column the RA in the basic interval of $\theta$ is given, while in
the second column the RA in the whole range of $\theta$ is given.
The quasiperiodic component is represented in gray dots and the
main periodic component is denoted by the solid line.}
\label{fig:ra1}
\end{figure}

\begin{figure}[p]
\caption{6th RA for $\eta=0.3$. The quasiperiodic component is
shown in gray, and the stable and unstable periodic orbits in the
gaps are denoted by solid and dashed lines, respectively. Main
period-$F_6$ and minor period-$3F_6$ components are shown
explicitly in (a) and (b), respectively.}
\label{fig:ra2}
\end{figure}

\begin{figure}[p]
\caption{Phase-dependent saddle-node bifurcation lines for the
main periodic components. The cases of the level $k=6,9,12$ 
are shown in (a), and other cases with $k=7,8,10$ are given in
(b). (c) Plot of $\ln{(1-\mu_k)}$ vs. $F_k$ for $\eta=0.3$. Solid
points denote the data for level $k=6,\dots,12$, which are
well fitted with a dashed straight line with slope
$\alpha=0.013$.} \label{fig:ra3}
\end{figure}

\begin{figure}[p]
\caption{Plot of $h^{(k)}_\varphi(\theta)$ vs. $\theta F_6$ for (a) $\eta=0.1$
and (b) $\eta=0.3$. (c) Plot of $<h^{(k)}_\varphi>$ vs. $\eta^2$ for
the three levels $k=6,7,$ and $8$. The solid and dashed lines
denote the Lyapunov exponents in the quasiperiodic limit that are
obtained numerically and analytically, respectively. }
\label{fig:ra4}
\end{figure}

\begin{figure}[p]
\caption{Attractors in two-dimensional surfaces. (a) $r$ versus $\theta$
at $\varphi = \pi$ surface and (b) $r$ versus $\varphi$ at $\theta=\pi$ with
$\rho=0.5$, $\lambda=0.5$, and $\eta=0.3$ ($h_\varphi = -0.024$ and $h_r = -1.370$).}
\label{fig:att}
\end{figure}

\begin{figure}[p]
\caption{Dimension computations for (\ref{eq:rdmap}) with $\eta = 0.3$,
$\lambda = 0.5$, $\rho = 0.5$,
$\omega=\pi(\sqrt{5}-1)$, $\bar{P}(\theta,\varphi)=\sin \varphi$,
and $Q(\theta,\varphi,r)=\sin(r+\varphi)$. In (a) the
dashed line has slope two, while in (b) it has slope one. In each graph,
the curves from lowest to highest represent $T=10^3,10^4,\cdots,10^{9}$;
in (b) the final four curves are virtually identical.}
\label{fig:bidr}
\end{figure}

\begin{figure}[p]
\caption{Construction of the domain $D_a$ (shaded region) of $g_a$.}
\label{fig:dia3}
\end{figure}

\begin{figure}[p]
\caption{Trajectory of the map (\ref{eq:3dmap}) with $\omega^{(1)}=\pi(\sqrt{5}
-1)$, $\omega^{(2)}=2\pi(\sqrt{2}-1)$, $\eta=0.3$,
$q^{(1)}=q^{(2)}=1$, and $P(\theta^{(1)},
\theta^{(2)},\varphi)=\sin (\theta^{(1)}+\varphi)$. 
In each case $\theta^{(1)}_0 = \theta^{(2)}_0
= \varphi_0 = 0$ and $10^4$ points of the trajectory are computed before
plotting the next $10^5$ points in the slice of (a) $\pi -\delta/2 < \theta^{(2)}
< \pi + \delta/2$ and (b) $\pi - \delta/2 < \theta^{(1)}< \pi+\delta/2$, 
where $\delta = 0.01$.}
\label{fig:traj3}
\end{figure}

\begin{figure}[p]
\caption{Dimension computations for (\ref{eq:3dmap}) with $\eta = 0.3$,
$\omega^{(1)}=\pi(\sqrt{5}-1)$, $\omega^{(2)}=2\pi(\sqrt{2}-1)$, and 
$P(\theta^{(1)},\theta^{(2)},\varphi)=\sin (\theta^{(1)}+\varphi)$. In (a) the
dashed line has slope three, while in (b) it has slope two. In each graph,
the curves from lowest to highest represent $T=10^4,10^5,\cdots,10^9$.}
\label{fig:bid3}
\end{figure}

\end{document}